\documentclass[useAMS,usenatbib]{mn2e}
\usepackage{color}
\usepackage{graphicx}
\usepackage{multirow}

\newcommand{\apj}{ApJ}

\newcommand{\mnras}{MNRAS}
\newcommand{\aap}{A\&A}

\newcommand{\nat}{Nature}

\newcommand{\pasj}{PASJ}

\title{The influence of the jet opening angle on the appearance of relativistic jets}
\author[T.~Boutelier et al.]
{T.~Boutelier, 
G.~Henri, 
P--O.~Petrucci 
\\ UJF-Grenoble 1 / CNRS-INSU, Institut de Plan\'etologie et dÕAstrophysique de Grenoble (IPAG) UMR 5274, Grenoble, F-38041, France.}
\date{Xxxxx XX}
\pagerange{\pageref{firstpage}--\pageref{lastpage}} \pubyear{2007}

\begin{document}
\label{firstpage}

\maketitle

\begin{abstract}
We reinvestigate the problem of the appearance of relativistic jets when geometrical opening is taken into account. We propose a new criterion to define apparent velocities and Doppler factors, which we think being determined by the brightest zone of the jet.  
We numerically compute the apparent velocity and the Doppler factor of a non homokinetic jet using different velocity profiles. We argue that if the motion is relativistic, the high superluminal velocities { $\beta_{app}\simeq \gamma$,  expected in the case of  an homokinetic jet, are only possible  for geometrical collimation smaller than the relativistic beaming angle $\gamma^{-1}$. {This is relatively independent of the jet velocity profile}. For jet collimation angles larger than $\gamma^{-1}$, the apparent image of the jet will always be dominated by parts of the jet traveling directly towards the observer {at lorentz factors $< \gamma$} resulting in maximal apparent velocities smaller than $\gamma$}. Furthermore, getting rid of the homokinetic hypothesis yields a complex relation between the observing angle and the Doppler factor, resulting in important consequences for the numerical computation of AGN population and unification scheme model.
\end{abstract}


\section{Introduction}\label{intro}
Jets are present in a very wide number of astrophysical objects, from the young stellar objects and X-ray binaries, at the galactic scale, to the powerful Active Galactic Nuclei { (hereafter AGN)} whose jets can cross hundreds of kpc in the intergalactic medium. If the jets are in relativistic motions, Doppler and time delay effects will greatly affect their appearance.  From a taxonomic point of view, the presence or absence of jets, and the inclination of the jet with the line of sight, is commonly used to classified extragalactic objects in different categories { (e.g. \citealt{urr95,urr03})} even if it is yet not clear that the jet characterizes a particular type of AGN or only a period of the cosmological evolution of galaxies.\\ 

Relativistic motion has been  proposed very early by \cite{ree66}  to solve the mystery of high luminosities and very rapid variability of extragalactic radio sources, which should have resulted in a catastrophic Inverse Compton cooling if the sources were at rest with respect to the observer. He predicted then the possibility of observing superluminal motions, a prediction beautifully confirmed thanks to the development of high resolution Very Large Base Interferometry. Later, detection of high energy gamma-ray radiation by EGRET led to the conclusion that relativistic motion was also necessary to escape the problem of self-absorption by pair production.  The most simple model that reproduces relatively well the observed { Spectral Energy Distribution (hereafter SED)} is the so-called one-zone Synchrotron self-Compton (SSC) model. It assumes a spherical zone in which high energy particles are injected and cooled via synchrotron and Inverse Compton processes. The model parameters are the zone radius, the magnetic field as well as the density and the distribution of the particles. To reproduce the observed SED however, a static model is generally ruled out due to causality constraints { (see \citealt{zen97} and references therein)}
and a necessarily low pair creation optical depth is needed in order to avoid  all gamma-rays photons being absorbed to form electron-positron pairs { (e.g. \citealt{mar92,hen93})}.  The solution is then to assume that the source is uniformly moving with a relativistic bulk velocity $v=\beta c$. In this case, all specific intensities, in the source rest frame, are a factor $\delta^3$ lower than in the observer frame, $\delta=1/\Gamma (1-\beta\mu)$ being the Doppler beaming factor, $\Gamma$ the usual Lorentz factor and $\mu=\cos\theta_{obs}$ the cosine angle of the jet axis with respect to the observer line of sight (LOS).  So the actual photon density in the jet frame is much lower than what would be deduced for a static source. The possible geometrical opening of the jet and non-uniformity of the Lorentz factor across the jet section are neglected in the simplest form of this model. \\

However jet opening angles are observed in different types of objects like  AGN or { Young Stellar Objects} \citep{jun99,hor06}. 
 Some of these observations show a decrease of the jet opening angle with distance from the central core, an indication of some collimation processes. Jet models also predict a variation of the jet opening angle (e.g. \citealt{fer97,cas02,mck06,haw06,zan07}) 
 and indeed some of them are able to reproduce the observations \citep{doug04}.\\

Given the strong dependance of the Doppler beaming factor on the angle between the line of sight and the direction of displacement of the emitting region, taking into account the  jet opening angle should have important impacts on the observed jet emission. In a series of papers,  Gopal-Krishna et al. (\citeyear{gop07a} and  reference therein. Hereafter GK) have investigated  some of these effects on the observed parameters of blazar jets like the jet orientation angle, its apparent speed and Doppler factor.

These authors proposed to compute the {\it effective} apparent velocity of the jet as a Doppler-boosted intensity weighted average of the apparent velocity of each point of the jet blob (cf Eq. (5) of \citealt{gop07a}). We think however that this prescription does not really reproduce the way by which apparent velocities are measured, i.e. by the displacement of the maximum of the image fit. Moreover, due to the light travel effects, neglected in the previous analyses, the shape of the emitting region is different  between the jet and the observer's frame, hence modifying the apparent shape on the sky plane. \\

In this paper we propose to re-analyse these different effects using our new prescription to estimate the apparent jet velocity. For sake of simplicity, we use a simple formalism for the jet geometry described in Sect. \ref{formalism}, and assume an infinitely thin shell propagating in the jet, which would capture the essential features of any perturbation traveling along the jet. We compare four different jet velocity profiles to see the influence of precise angular distribution of Lorentz factors. We detail the way we compute the apparent velocity of the whole pattern seen by an observer at infinity in Sect. \ref{Appvel}. We then present our results in Sect. \ref{results} and discuss them in Sect. \ref{sec:disc} before giving concluding remarks in Sect. \ref{sec:conc}.

%
\section{The model}\label{sec:model}
\subsection{Formalism}
\label{formalism}
\begin{figure}
\includegraphics [width=\linewidth]{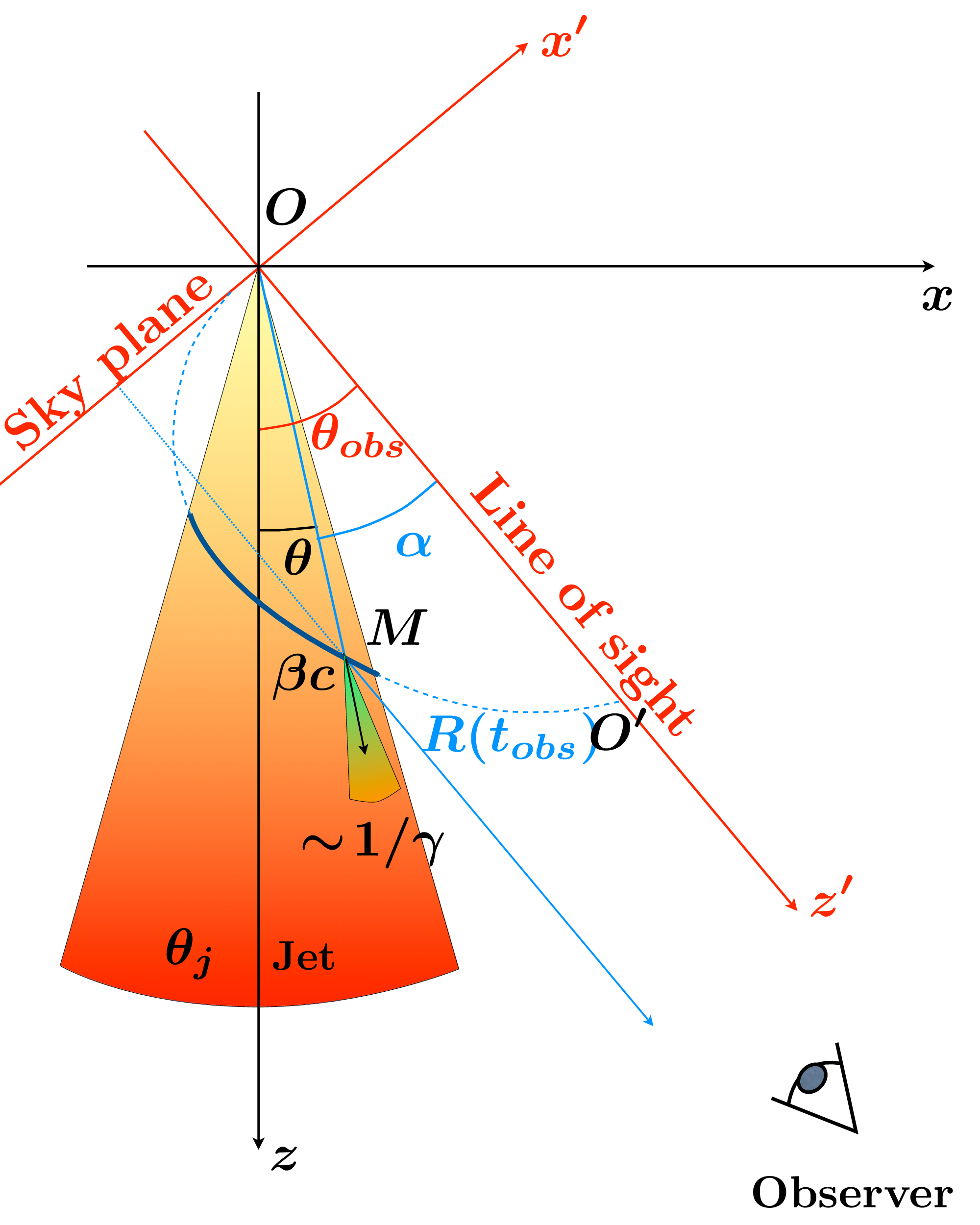}
\caption{Sketch of the jet model in the case of velocity distribution $D_{1}$. See text for the signification of the different parameters.}
\label{fig:model-scheme}
\end{figure}
We consider, in the following work, the simple case of a instantaneous central perturbation,   propagating with a relativistic speed characterized by the Lorentz factor $\gamma_{0}$ on the jet axis. In the jet rest frame, we assume that the surface emissivity is uniform, its spectrum being described by a power-law of index $n$ defined by $S_{\nu}\propto\nu^{-n}$ { (see Sect. \ref{Iangdep} for the case of an angular dependent emissivity)}. The geometrical collimation of the jet is characterized by a characteristic parameter $\theta_{j}$, whose exact definition depends on the assumed shape. The velocity distribution is described by a function $\gamma(\theta)$, where $\theta$ is the angle to the jet axis. For a point of this surface at a given polar angle $\theta$, the velocity vector also { points in the radial} direction { (see Fig. \ref{fig:model-scheme})}.\\ 

Four different velocity profiles will be used to study the influence of the jet structure on the observational parameters, and compared with the single blob, homokinetic results. The first one ($D_{1}$) considers the simple case of a conical surface with a constant Lorentz factor $\gamma_{0}$ inside a cone of half-opening angle $\theta_{j}$, and a Lorentz factor equal to unity outside (i.e. null velocity). This distribution is obviously not physical due to the sharp discontinuity at the edge of the cone, but is convenient to use. It is the same as the one used in most gamma-ray bursts models. In the absence of any specific form argued for in the literature, we have also studied different, smoother distributions. The second distribution ($D_{2}$) is built using the conical distribution $D_{1}$, and adding a power-law decrease of index $-2$  for the Lorentz factor outside the cone. The two last velocity distributions are gaussian ($D_{3}$) and lorentzian ($D_{4}$) profiles. Then for every $\left| \theta \right| < \pi/2$ { (approaching jet)}, we define:
\begin{eqnarray}
  D_{1}:\gamma (\theta)&=&\left\{ 
    \begin{array}{ll} 
      \gamma_{0} & \rmn{ if } \left| \theta \right| <\theta_{\rmn{j}} \label{eq:D1}\\
      1 & \rmn{ else } 
    \end{array} \right.\\ 
  D_{2}:\gamma (\theta)&=&\left\{
    \begin{array}{ll} 
      \gamma_{0} & \rmn{ if } \left| \theta \right| <\theta_{\rmn{j}}\\
      1+(\gamma_{0}-1)\left(\displaystyle\frac{\theta}{\theta_{\rmn{j}}}\right)^{-2}  & \rmn{ else } 
    \end{array} \right.  \label{eq:D2}\\
 D_{3}:\gamma (\theta)&=& 1+(\gamma_{0}-1)\exp
  \left[-\ln2\left(\displaystyle\frac{\theta}{\theta_{\rmn{j}}}\right)^{2}\right] \label{eq:D3} \\
  D_{4}:\gamma (\theta)&=&
  1+\frac{(\gamma_{0}-1)}{1+\left(\displaystyle\frac{\theta}{\theta_{\rmn{j}}}\right)^{2}}  \label{eq:D4}
\end{eqnarray}
The symmetrical situation holds for $\left| \theta \right| > \pi/2$ (counter-jet): $\gamma(\theta)=\gamma(\theta+\pi)$.\\

For a given  observation angle $\theta_{\rmn{obs}}$ defined between the jet axis and the line of sight, we project on the sky plane the surface observed at a given observational time $T_{\rmn{obs}}$. The observed flux on the sky plane is related to the intrinsic flux in the source rest frame $S_{\nu,\,\rmn{int}}$ by the Doppler factor: $S_{\nu,\,\rmn{obs}}=S_{\nu,\,\rmn{int}}\delta^{3+n}$. Due to the velocity distribution $\gamma(\theta)$, each point shell have an intrinsic velocity different in norm and direction, and then a different apparent speed as measured by the observer.

The appearance of a moving feature must also be corrected by time delay effects to the observer.  Let us define two coordinate frames : $\mathcal{R}_{1}$ ($Ozx$ on Fig.~\ref{fig:model-scheme}), for which one of the axis is aligned with the jet axis, and $\mathcal{R}_{2}$ ($Oz'x'$ on Fig.~\ref{fig:model-scheme}), aligned with the observer line of sight, and obtained by rotating $\mathcal{R}_{1}$ by an angle $\theta_{\rmn{obs}}$. If one defines a point $M$ with the polar coordinates ($r,\theta$) in $\mathcal{R}_{1}$, hence the polar coordinate of $M$ in $\mathcal{R}_{2}$ are ($r,\alpha=\theta-\theta_{\rmn{obs}}$). The observation time of $M$, i.e. the instant when the light emitted at $M$ (when the shell just reached it) will be seen by the observer is:
\begin{equation}
T_{\rmn{obs}}(M)=\frac{r}{\beta(\alpha) c} - \frac{r\cos\alpha}{c} = \frac{r}{\beta(\alpha) c}(1-\beta(\alpha)\cos\alpha)
\label{eqtobs}
\end{equation}
where $\beta(\alpha)$ is deduced from the velocity distribution $\gamma(\theta)$ expressed in $\mathcal{R}_{2}$:
\begin{equation}
\beta(\alpha) = \sqrt{1-\frac{1}{\displaystyle\gamma(\theta=\alpha+\theta_{\rmn{obs}})^2}}
\end{equation}
Hence, two points of the jet  $M_{1}$ and $M_{2}$ will be seen by the observer at the same instant if the observation times reach the condition: $T_{\rmn{obs}}(M_{1})=T_{\rmn{obs}}(M_{2})$. Let's choose as a reference point the intersection between the propagating shell and the jet axis. At a given instant $t$, this point is at distance $r_{0}(t)$ from the origin, and is characterized by the Lorentz factor $\gamma_{0}$. Using Eq. (\ref{eqtobs}), the parametric equation of the jet surface seen at a given observational time expressed in $\mathcal{R}_{2}$ is then:
\begin{equation}\label{eq:TobsCt}
r(\alpha)=r_{0}(t) \left(\frac{\beta(\alpha)}{\beta_{0}}\right) \left[ \displaystyle \frac{1-\beta_{0}\cos\theta_{\rmn{obs}}}{1-\beta(\alpha)\cos\alpha} \right]
\end{equation}
\begin{figure}
\includegraphics [width=\linewidth]{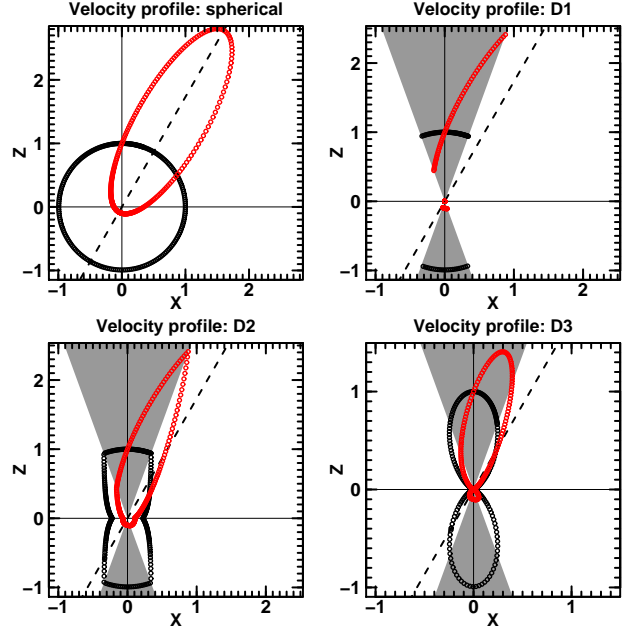}
\caption{Surfaces seen by an observer at a given arbitrary observation time $T_{\rmn{obs}}$ (Eq.~\ref{eq:TobsCt} with $r_{0}=1$), for the non-relativistic case ($\gamma_{0}= 1.00001$, in black), and for the relativistic case ($\gamma_{0}=3$, in red), for a spherical expanding surface (Upper-left case), and the different velocity distributions $D_{1-3}$. The shaded region represents the geometrical collimation of the jet ($\theta_{j}=20\,^\circ$), and the dashed line is the observer line of sight ($\theta_{\rmn obs}=30^\circ$).}
\label{fig:Surface}
\end{figure}
No characteristic scale is involved in this equation which is self-similar. Thus, the observed surface at a different observational time is a simple homothetic transformation of the previous one, through the time dependent evolution of $r_{0}$. In the non-relativistic case ($\beta \ll 1$), one can easily check that this expression simply yields  the intrinsic shape of the surface. Figure~\ref{fig:Surface} shows the deformation of the observed surface for a purely spherical velocity profile and the 3 different velocity profiles $D_1$, $D_2$ and $D_3$, using $\gamma_{0}=3$. Even for this mildly relativistic regime, the deformation due to light propagation delay is strong. Hence, this effect must be taken into account if one wants to compute properly the observed intensity, projected on the sky plane. For comparison, the shape of the shell in a non-relativistic case ($\gamma_0 -1 = 10^{-5}$) is shown. In this case, as we remarked,  the shape is the real one since the light travel effects are negligible. \\

\begin{figure*}
\vspace*{-1.5cm}
\includegraphics [height=0.8\textwidth,angle=90]{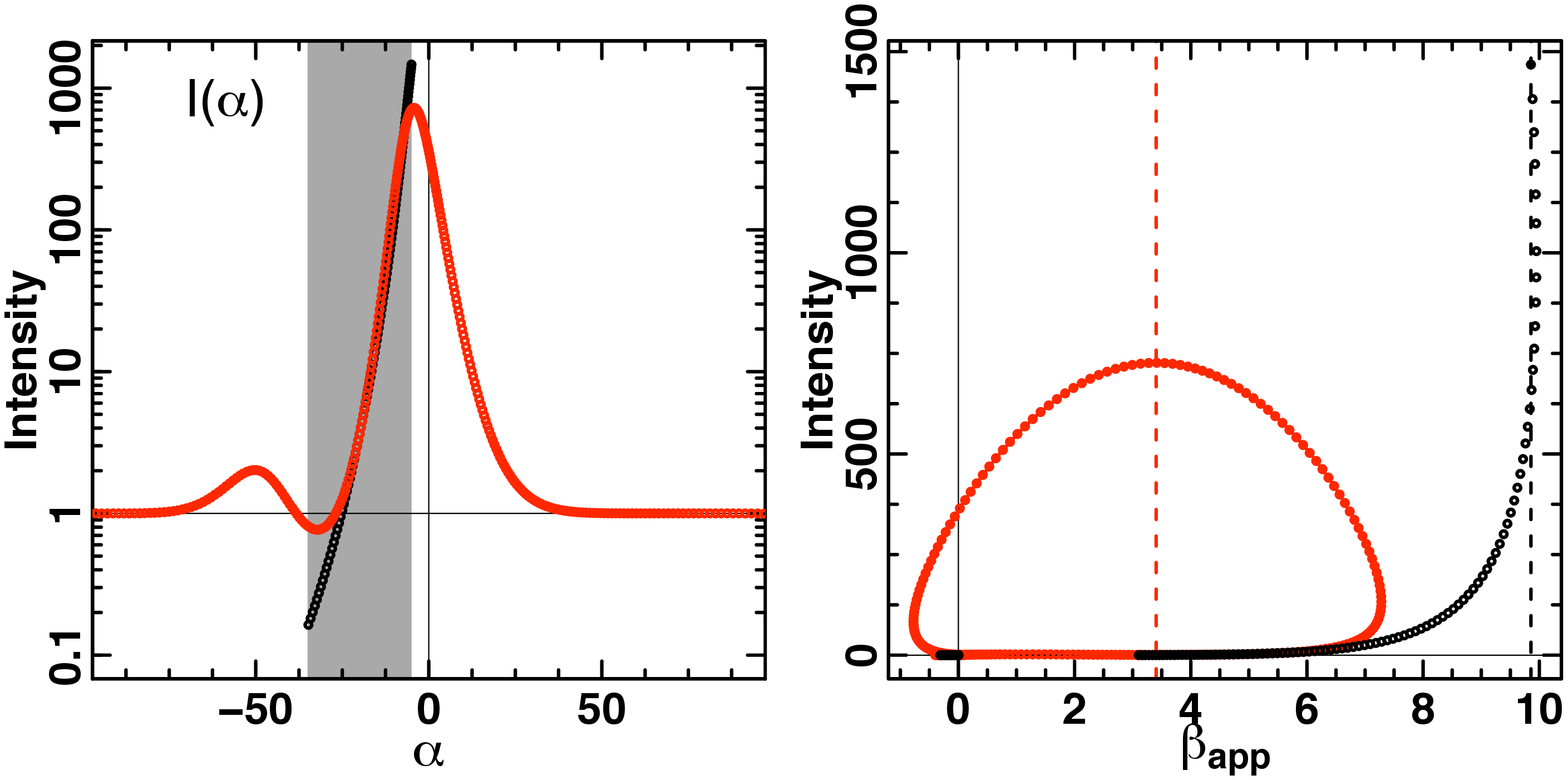}
\vspace*{-1.5cm}\caption{Comparison of the jet apparent velocity for the conical and gaussian profile $D_1$ and $D_3$ and for an observation angle $\theta_{obs}=20^{\circ}$, a jet opening angle $\theta_{j}=15^{\circ}$ and a Lorentz factor on the jet axis $\gamma_0=10$ . 
{\bf Left:} Plots of the two intensity profiles $I(\alpha)$ (black: conical profile, red: gaussian profile), computed following Eq. (\ref{eqI}). Their maxima are not at the same position compared to the line of sight (vertical solid line) and are not obtained for the same value of the Lorentz factor. { The gray area corresponds to the geometric jet aperture i.e. $\mathbf{|\alpha -\theta_{obs}|\le\theta_j}$}.{\bf Right:} Plot of the same intensity profiles but versus the apparent speed $\beta_{app}$ of each point of the shell surface. The apparent speed of the whole pattern is then given by the apparent speed of the brightest point of the shell, indicated by the vertical dashed lines. In the present example, this apparent speed is equal to 9.85c and 3.4c for the conical and gaussian case respectively.}
\label{fig:projection}
\end{figure*}

Knowing the shape of the observed surface, it is now possible to project it on the sky plane and to compute the intensity profile at each point of the sky. A point $M$ of this surface is characterized by its polar coordinate $\alpha$ in  $\mathcal{R}_{2}$, and its radius $r(\alpha)$ given by Eq.~(\ref{eq:TobsCt}). The projection of $M$ on the sky plane is simply given by:
\begin{equation}
x'(\alpha)=r(\alpha)\sin\alpha
\end{equation}

The apparent velocity (in unit of $c$) of $M$ is computed using the formula for a relativistic point source:
\begin{equation}\label{eq:BetaApp}
\beta_{\rmn{app}}(\alpha)=\frac{\beta(\alpha)\sin\alpha}{1-\beta(\alpha)\cos\alpha} = \frac{x'(\alpha)}{r_{0}} \left( \frac{\beta_{0}}{1-\beta_{0}\cos\theta_{\rmn{obs}}} \right)
\end{equation}
It is interesting to note that, in our formalism, the apparent speed can be expressed in function of the position on the sky plane due to the self-similarity of the observed surface. The Doppler factor of $M$ is:
\begin{equation}
\delta(\alpha)= \frac{\sqrt{1-\beta(\alpha)^2}}{1-\beta(\alpha)\cos\alpha }
\end{equation} 
Assuming a flat spectrum source ($n=0$) and a uniform intrinsic emissivity, the observed intensity is simply $I(\alpha)=I_{0}\delta(\alpha)^3$. Then the intensity profile is parametrized by these two equations:
\begin{equation}
\label{eqI}
\left\{
\begin{array}{l}
x'(\alpha)=\displaystyle r_{0}\sin\alpha \left(\frac{\beta(\alpha)}{\beta_{0}}\right) \left[ \displaystyle \frac{1-\beta_{0}\cos\theta_{\rmn{obs}}}{1-\beta(\alpha)\cos\alpha} \right] \\
I(\alpha)= \displaystyle I_{0} \left[ \frac{\sqrt{1-\beta(\alpha)^2}}{1-\beta(\alpha)\cos\alpha } \right]^3
\end{array} 
\right. 
\end{equation}

\subsection{Apparent velocity of a shell}\label{Appvel}
\begin{figure*}
\begin{tabular}{cc}
\includegraphics[width=0.4\linewidth]{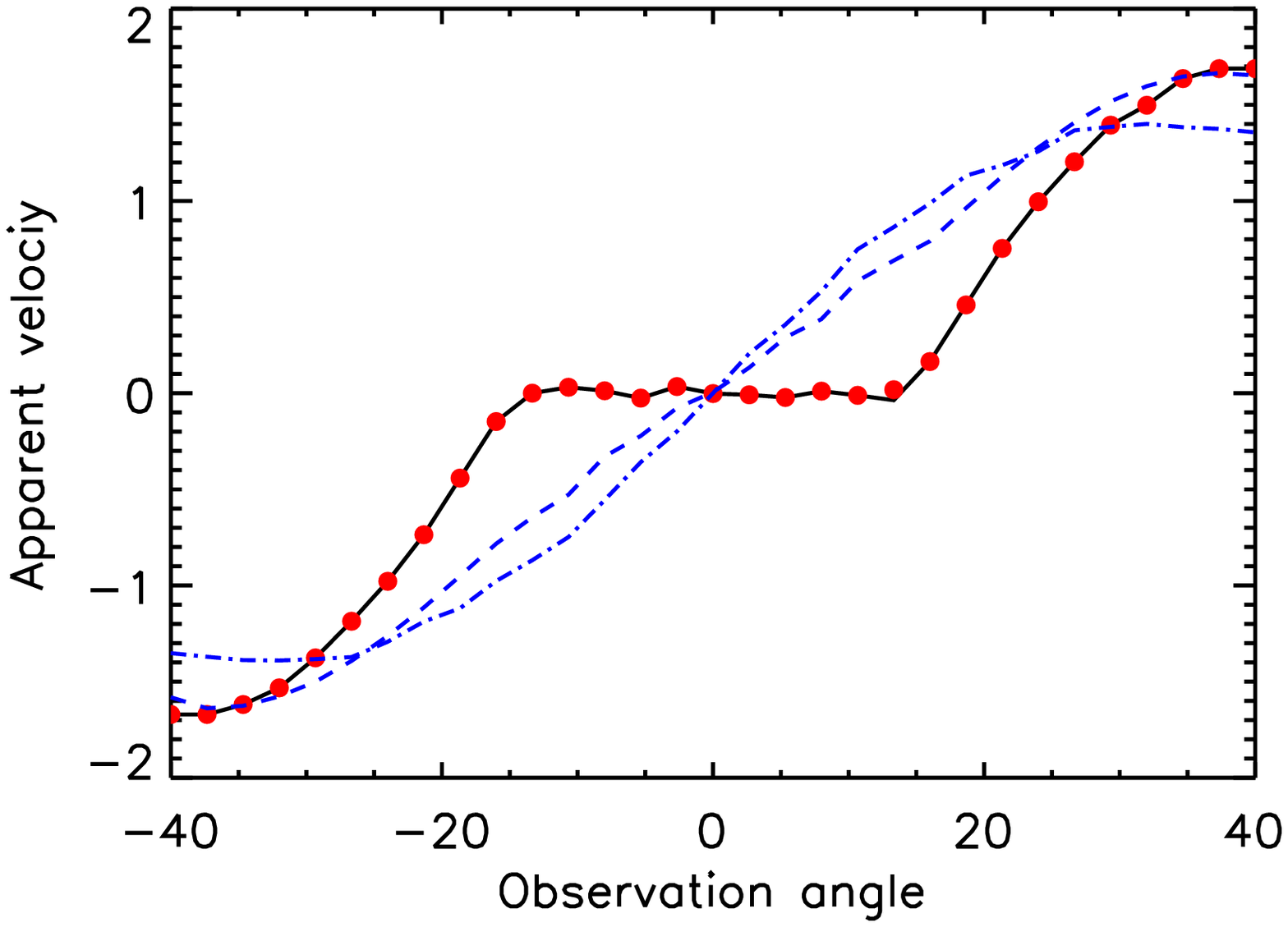}&
\includegraphics[width=0.4\linewidth]{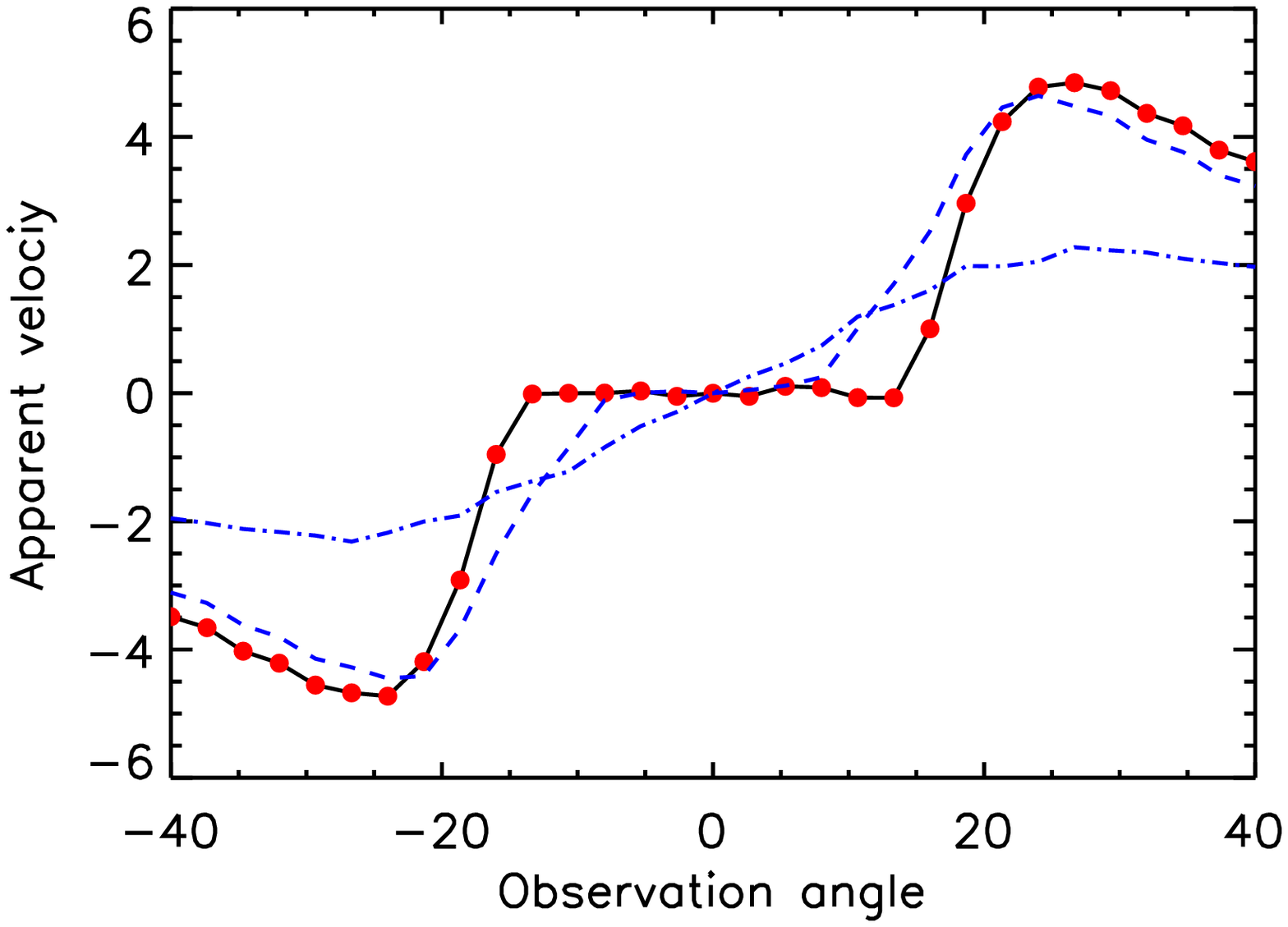}
\end{tabular}
\caption{\label{fig:appvelocity} Comparison of four different prescriptions for the apparent velocity of a thin shell. The left panel is computed with $\gamma_0= 2$, and the right panel with  $\gamma_0= 5$. Solid line: this paper's prescription. Dot-dashed line: original GK07's prescription with an averaged velocity, without light-travel effects; Dashed line: modified prescription of GK07 including light travel effects.  Red dots: results of the gaussian fit with a simulated image. 
}
\end{figure*}

The above formulae hold for a single emission location. However, in real observations, one deals with a complete pattern of emission and an observer will derive an apparent velocity of the whole pattern. How this apparent velocity is defined is not a simple issue and may depend on how the different components are identified and followed. \citet{gop07a} have adopted a prescription by computing an effective apparent velocity as a Doppler-boosted intensity weighted average of the apparent velocity of each point of the emitting pattern (cf Eq. (5) of \citealt{gop07a}). We think however that this prescription does not really reproduce the way by which apparent velocities are derived. Usually, observers will fit the components by a bell-shape (often gaussian) try function (e.g. \citealt{lis09}) and define the velocity by the displacement of the maximum of the fit. Thus we have chosen another prescription: we define the effective apparent velocity of a component as the velocity of the maximum of specific intensity, much like one defines the group velocity of a wave packet.  \\
{
The difference can be illustrated in a simple case, where a conical shell is expanding at a constant velocity. If the observer line of sight lies within the cone of expansion, but not straight on-axis, he will see an asymmetrical pattern, but whose maximum will always be at zero angle with the line of sight, since the Doppler factor will always be maximal in this direction. An average velocity will thus be non-zero, whereas the maximum will not move.
 To illustrate the difference with GK07's prescription, we have simulated a conical jet with a constant Lorentz factors and a finite opening angle $\theta_j = 15 \deg $. The Lorentz factor has been fixed to $\gamma_b = 2 $ or  $\gamma_b = 5 $. We note that  GK07 didn't take into account the apparent deformation of the surface caused by light travel effects, and have computed the averaged velocity for a given proper time and not a constant observer time. We thus computed the apparent velocity in four different ways : taking the initial GK07's prescription without light-travel time correction, a modified GK07's prescription taking into account the light travel time, our prescription (velocity of the brightest point), and the velocity deduced from a gaussian fit of the intensity map (with light travel time corrections). The results are displayed on Fig. \ref{fig:appvelocity}. We see that our prescription with the brightest point is in complete agreement with what would be inferred from a gaussian fit of the intensity map, then supporting our own prescription. Inclusion of light travel effects in the GK07 prescription produces significant differences with their original calculations, especially for large Lorentz factor (right panel of Fig. \ref{fig:appvelocity}). Interestingly, this "modified" prescription looks rather similar to our results.}\\

Due to the axial symmetry hypothesis on the jet geometry, the problem of determining the position of the brightest point of a tridimensional surface projected on the sky plane can be treated in a bi-dimensional approach. Indeed, the maximum of intensity is necessarily in the plane defined by the jet axis and the observer line of sight, i.e. the plane of Fig.~\ref{fig:model-scheme}. The position of the maximum is computed numerically, using a ``Golden Section search'' algorithm \citep{Kiefer53}. 
As an example, Fig. \ref{fig:projection} shows a "visual" comparison of the different steps of the computation of the jet apparent speed for the conical and gaussian profile $D_1$ and $D_3$. 
The two intensity profiles $I(\alpha)$, computed following Eq. (\ref{eqI}), are plotted in the left panel of Fig. \ref{fig:projection}. Their maxima are not at the same position compared to the line of sight and are not obtained for the same value of the Lorentz factor. This results in a different apparent velocity of the whole pattern on the sky plane as shown in the right panel of Fig. \ref{fig:projection}. In consequence the velocities of the maximum of intensity, which correspond to the apparent velocities of the jet in our formalism, are different of the order of 9.85c for the $D_1$ profile, and 3.4c for the $D_2$ profile.

\begin{figure}
\includegraphics [height=50mm]{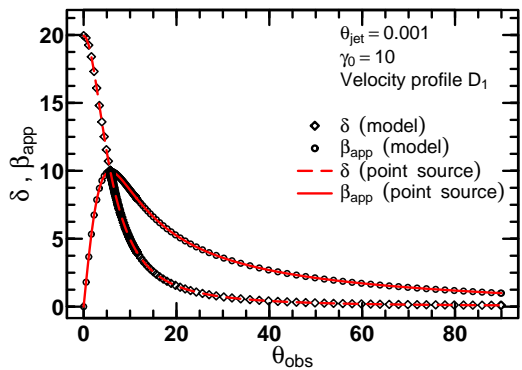}
\caption{Apparent velocity in unit of $c$ (black open circle), and Doppler factor (black open diamond), as a function of $\theta_{\rmn{obs}}$, computed with our conical jet model (velocity profile $D_{1}) $ assuming $\theta_{j}=0.001\,^\circ$ and $\gamma_{0}=10$. We also display the theoretical curves for the apparent velocity (red line) and Doppler factor (red dotted line) of a point source of same Lorentz factor $\gamma_{0}=10$. The agreement is excellent.}
\label{fig:BetaAppTest}
\end{figure}

\subsection{Test case: perfectly collimated jet}
We have tested our method to compute the appearance of a relativistic jets (apparent velocity and Doppler factor) in the case of a very narrow collimated jet. The geometrical angle being much smaller than the radiation cone, the result should be very close to the point source case. The agreement is indeed excellent as shown in  Fig. \ref{fig:BetaAppTest}, our model of very well collimated jet mimicking perfectly the behavior of a point like source.\\


%
\section{Results}
\label{results}
We are now able to compute the apparent velocity of a jet with a given velocity profile ($\theta_{j},\,\gamma_{0},\,D_{i}$), and for a given observation angle $\theta_{\rmn{obs}}$. By iterating the process for different $\theta_{\rmn{obs}}$, we are able to compute the relation between the observation angle and the apparent speed of the whole pattern, and to determine the  maximum apparent velocity that can be reached for a given jet configuration, as well as the associated Doppler factor, for all possible orientations.

\subsection{Effect of the jet velocity profile}
\begin{figure*}
\includegraphics[width=0.4\linewidth]{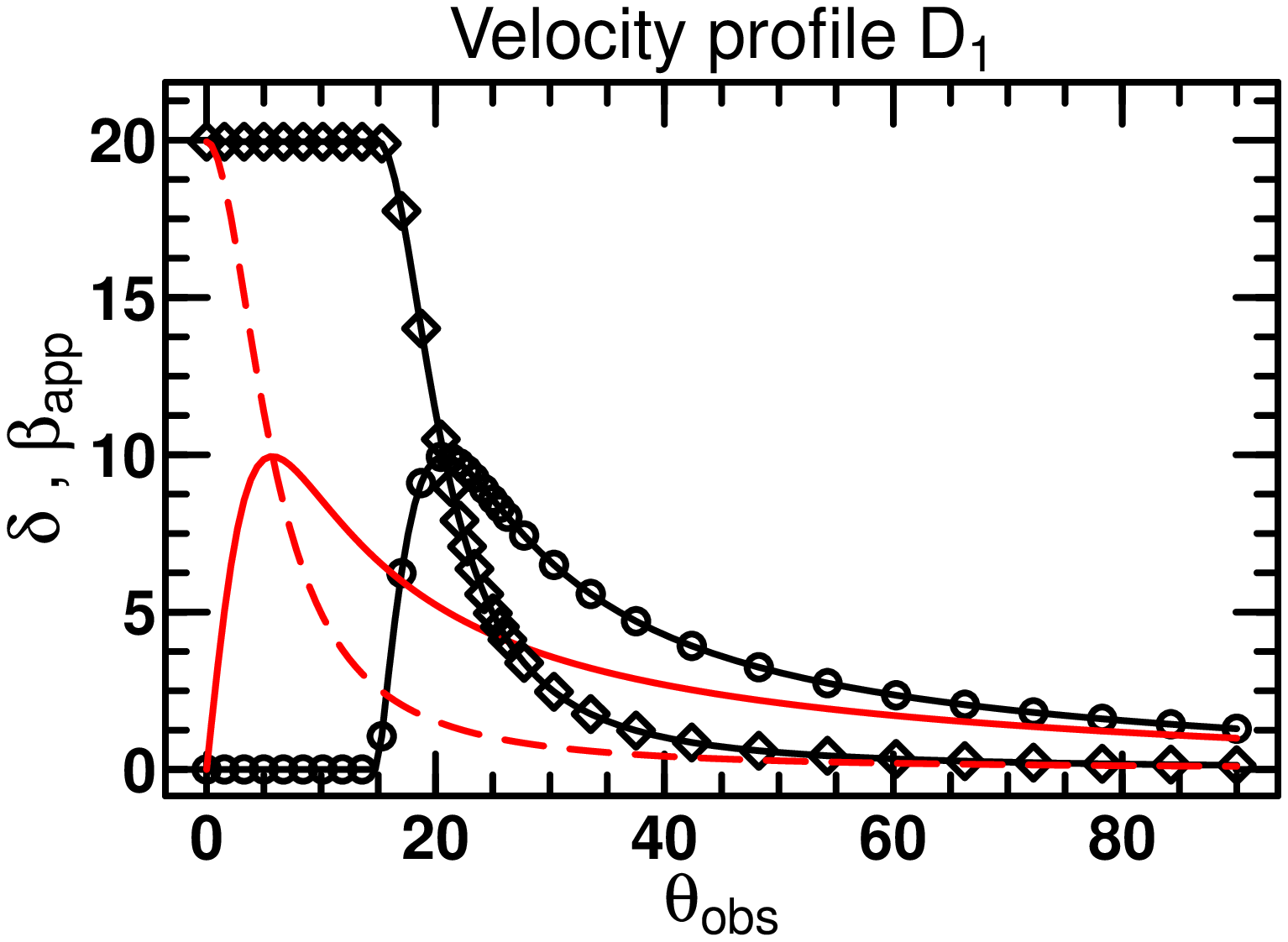}
\includegraphics[width=0.4\linewidth]{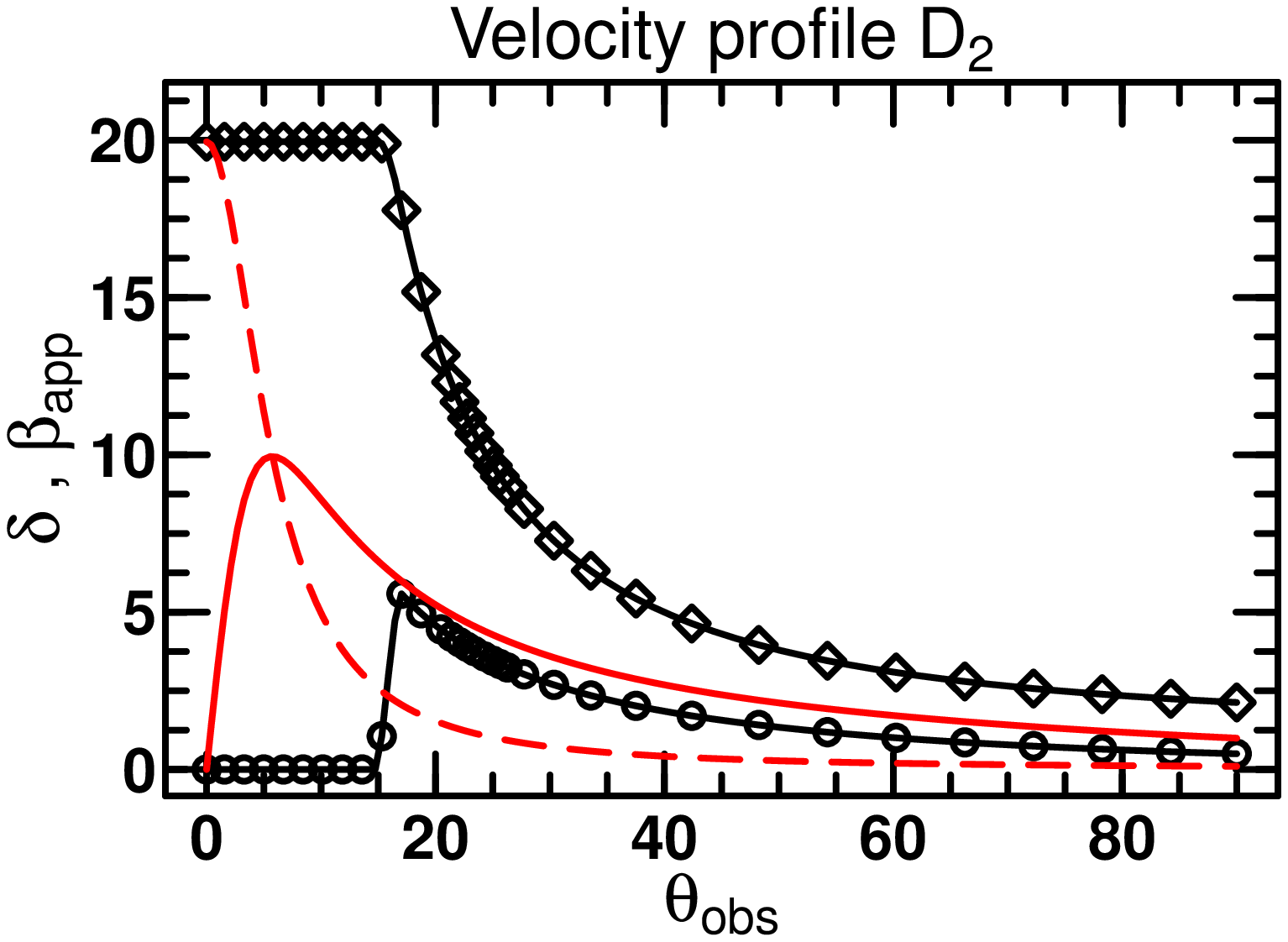}
\includegraphics[width=0.4\linewidth]{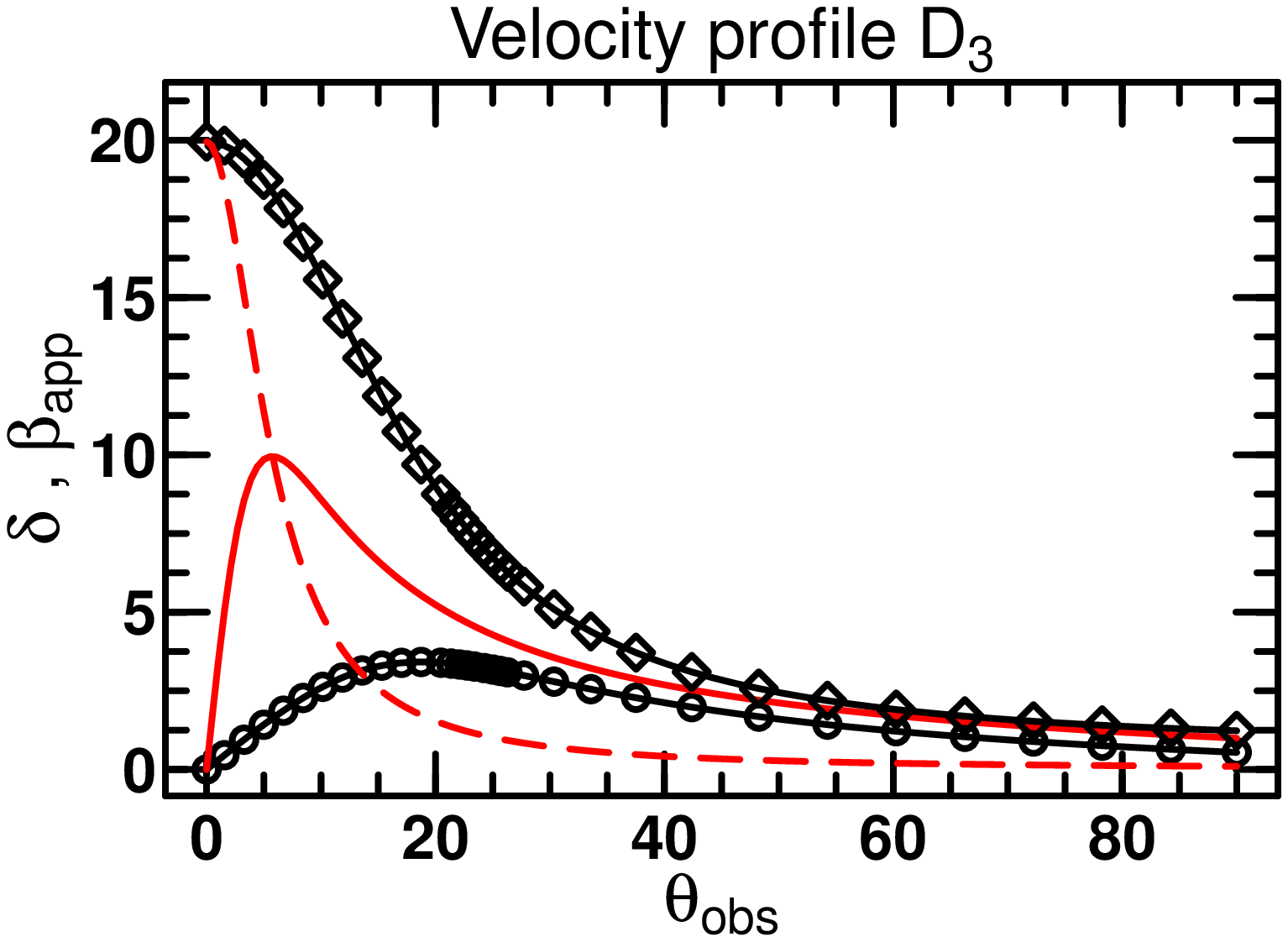}
\includegraphics[width=0.4\linewidth]{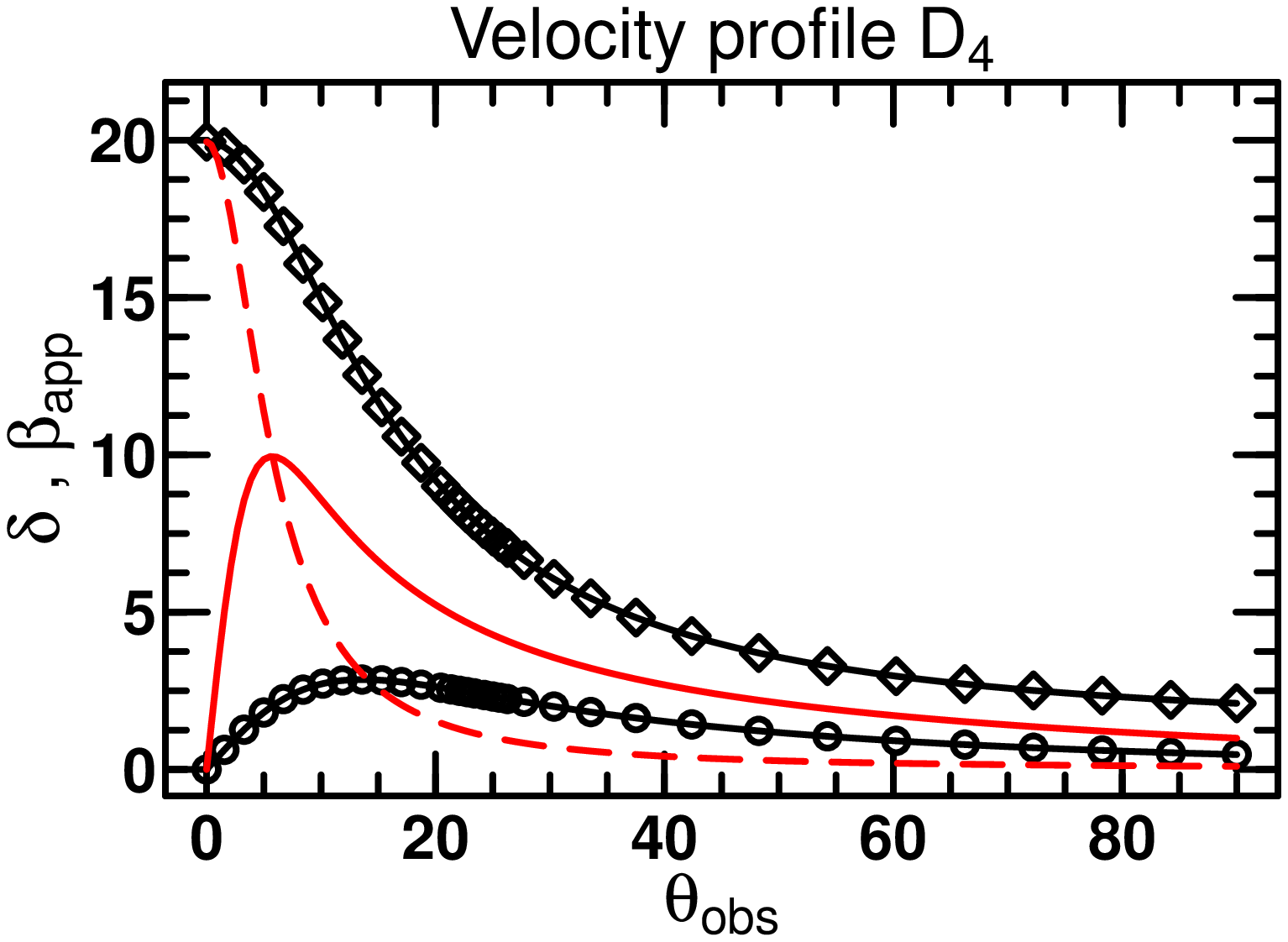}
\caption{\label{fig:BetaAppProfil}Jet apparent velocity (black empty circles) in unit of $c$ and Doppler factor (black empty diamonds), as a function of the observation angle $\theta_{\rmn{obs}}$, computed by our model for different jet velocity profiles. Upper-left corner: velocity distribution $D_{1}$ (Eq.~\ref{eq:D1}), upper-right corner:  velocity distribution $D_{2}$ (Eq.~\ref{eq:D2}), lower-left corner:  velocity distribution $D_{3}$ (Eq.~\ref{eq:D3}), lower-right corner:  velocity distribution $D_{4}$ (Eq.~\ref{eq:D4}). The jet opening angle is $\theta_{j}=15\,^\circ$, and the Lorentz factor on the jet axis is $\gamma_{0}=10$. For comparison, we have also plotted the theoretical curves for the apparent velocity (red line) and the Doppler factor (red dashed line) of a point source with a Lorentz factor of 10.
}
\end{figure*}

Figure~\ref{fig:BetaAppProfil} shows the evolution of the jet apparent velocity and the Doppler factor as a function of the observation angle $\theta_{\rmn{obs}}$ for the four velocity profiles $D_{1-4}$ defined in Eqs.~\ref{eq:D1}$-$\ref{eq:D4}. We assume a geometrical collimation $\theta_{j}=15\,^\circ$ and a Lorentz factor on the jet axis fixed to $\gamma_{0}=10$. As a comparison, we also display the theoretical curves of the apparent velocity (red line) and Doppler factor (red dashed line) expected in the case of a perfectly collimated jet  with the same Lorentz factor.   


%
\subsubsection{Conical jet}
\label{con_jet}
In the case of the conical velocity profile $D_{1}$ (upper-left corner of Fig.~\ref{fig:BetaAppProfil}), and as we noted previously,  the apparent velocity is equal to 0 for $\theta_{\rmn{obs}}\leq\theta_{j}$, while the Doppler factor is maximal ($\delta=2\gamma_{0}$). For a higher observation angle, the apparent velocity and the Doppler factor follow the same behavior as the theoretical expressions, but shifted by an angle $\theta_{j}$.\\ 

These results can be understood rather easily: while the observer is looking inside the jet cone ($\theta_{\rmn{obs}}\leq\theta_{j}$), the observed image is dominated by the part of the shell that points directly toward the observer. Then, the measured apparent velocity is null, and the Doppler factor is maximal ($2\gamma_{0}$). In this case, the image seen by the observer will be a distorted circle, expanding with time homothetically around the brightest point of the image. 

As soon as the line of sight is out of the cone ($\theta_{\rmn{obs}}>\theta_{j}$), the image is dominated by the emission coming from the edge of the conical jet. Then, the situation becomes similar to a point-like source, seen under the angle $\theta_{\rmn{obs}}-\theta_{j}$. The evolution of the apparent velocity $\beta_{\rmn{app}}(\theta_{\rmn{obs}})$ and the Doppler factor $\delta (\theta_{\rmn{obs}})$ is then the same as for the theoretical one in case of a perfectly collimated jet, shifted by $\theta_{j}$, the external ridge playing the role of a thin narrow jet.\\ 

For this kind of velocity profile, it is always possible to observe high super-luminal motion if the jet is seen close to the edge of the cone. For each cone angle $\theta_{j}$, the maximal apparent speed reachable is  $\beta_{\rmn{app,\,max}}\approx\gamma_{0}$ for an observation angle $\theta_{\rmn{obs}}\approx \theta_{j}+\gamma_{0}^{-1}$. The Doppler factor is then $\delta\approx\gamma_{0}$. But more interestingly, a wide range of observation angles can be associated to an absence of apparent motion, while the Doppler factor is maximum. This situation is unlikely if one consider a perfectly collimated jet, but become highly probable if the jet opening angle is large.

%
\subsubsection{Jet with a continuous velocity profile}
Profile $D_{2}$ is obtained from $D_{1}$ by adding a power-law decrease of the Lorentz factor outside the jet cone (Eq.~\ref{eq:D2}), so that the jet velocity smoothly decreases to zero at large $\theta$. The corresponding curves of the shell apparent velocity and Doppler factor as a function of the observation angle (upper-right corner of Fig.~\ref{fig:BetaAppProfil}) are the result of a complicated convolution of the Lorentz factor profile, with the theoretical formula for a point-source. Similar to the $D_{1}$ profile, the apparent velocity is null when the observer looks inside the cone ($\theta_{\rmn{obs}}<\theta_{j}$). As soon as the observation angle is larger than the collimation angle ($\theta_{\rmn{obs}}>\theta_{j}$), the brightest point of the jet moves out from the line of sight, yielding to a non-zero apparent velocity. However, for a given observation angle $\theta_{obs}$, the apparent velocity is lower than in the case of $D_{1}$. This is due to 2 reasons: 1) the lower intrinsic Lorentz factor $\gamma$ at the position of the brightest point on the shell, compared to the Lorentz factor value $\gamma_0$ on the jet axis, and 2) this brightest point is very close to the line of sight with $\alpha<\gamma^{-1}$. This last point is confirmed by the high value of the Doppler factor ($\delta>\beta_{\rmn{app}}$). \\

The lower part of Fig.~\ref{fig:BetaAppProfil} shows the results of the model for a gaussian velocity profile ($D_{3}$, Eq.~\ref{eq:D3}) and a lorentzian profile ($D_{4}$, Eq.~\ref{eq:D4}). For both distributions the apparent velocity is never null for any observation angle but $\theta_{\rmn{obs}}=0\,^\circ$, the brightest point of the shell being always slightly shifted out the line of sight. However, we emphasize that the apparent velocity is significantly smaller compared to the point-like source, for the same two reasons explained before. As for the $D_{2}$ velocity profile, this is confirmed by the high values of Doppler factor: $\delta>\beta_{\rmn{app}},~\forall~\theta_{\rmn{obs}}$. We can observe in Fig.~\ref{fig:BetaAppProfil} that the maximal apparent speed is reached for an observation angle close to the jet opening angle $\theta_{\rmn{obs}}\approx \theta_{j}\pm \epsilon$.

\subsubsection{Analytical approximation for the apparent velocity}
The previous analysis seems to indicate that the maximal apparent speed that can be observed with a given velocity profile is linked to the angular velocity gradient inside the jet. Indeed, $\beta_{\rmn {app}}$ is maximal with $D_{1}$, for which the velocity discontinuity at the edge of the cone induces a large gradient. The lowest  $\beta_{\rmn {app}}$ are obtained with the Lorentzian profile ($D_{4}$) which is the smoothest profile { (for angles greater than $\theta_j$)} among the four. Furthermore, $\beta_{\rmn{app}}$ is null when the observer points towards a constant velocity region in the jet (see the case of $D_{1}$ and $D_{2}$ with $\theta_{\rmn{obs}}<\theta_{j}$, upper part of Fig.~\ref{fig:BetaAppProfil}). The exact analytical study of the expression which gives the apparent velocity $\beta_{\rmn{app}}(\theta_{\rmn{obs}})$ is pretty involved. However, a Taylor development at the first order of the Doppler factor as a function of the angle $\alpha$ in $\mathcal{R}_{2}$ can explain this general behavior.

Indeed, for high Lorentz factors ($\gamma \gg 1$) and small observation angles ($\alpha<\gamma^{-1}$), the Doppler factor can be written as:
\begin{equation}
\delta(\alpha,\gamma)\approx \frac{2\gamma}{1+\gamma^2\alpha^2}
\end{equation}
In our model, we make the assumption that the intrinsic jet emissivity is constant. Hence, the position of maximum observed intensity corresponds to the position of the maximum Doppler factor. Thus the logarithmic derivative of the previous equation vanishes at the position of maximum observed intensity:
\begin{equation}
\frac{\rmn{d} \delta}{\delta} = \frac{\rmn{d} \gamma}{\gamma} - \frac{2\gamma\alpha^2\rmn{d}\gamma + 2\alpha\gamma^2\rmn{d}\alpha}{1+\gamma^2\alpha^2}  = 0
\end{equation} 
After some calculation, this equation can be reduced to a simple quadratic equation for $\alpha$:
\begin{equation}
\alpha^2 + 2\frac{\gamma}{\dot{\gamma}}\alpha  - \frac{1}{\gamma^2}=0 \quad\textrm{with}\quad \dot{\gamma}=\frac{\rmn{d}\gamma}{\rmn{d}\alpha}
\end{equation}
The two solutions of this equation are:
\begin{equation}
\alpha_{\pm} = -\frac{\gamma}{\dot{\gamma}} \left[ 1 \pm \sqrt{1 + \frac{\dot{\gamma}^2}{\gamma^4}}  \right] 
\end{equation}

We will consider two cases, following the sharpness of the velocity profile. A smooth profile will be defined by $\left |\dot{\gamma}\right |\ll\gamma^2$, meaning that the decrease of $\gamma$ takes place over an angle interval $\displaystyle\Delta \alpha \simeq \frac{\gamma}{\left | \dot{\gamma}\right |} \gg \frac{1}{\gamma}$. In this case, the square root in the previous expression can be linearized and we obtain the solutions:
\begin{equation}
\alpha_{+} \approx - \left( \frac{2\gamma}{\dot{\gamma}} + \frac{\dot{\gamma}}{2\gamma^3} \right) \quad\textrm{and}\quad \alpha_{-}\approx \frac{\dot{\gamma}}{2\gamma^3}\label{eq:Alphapm}
\end{equation}
We can easily check that the first solution corresponds to a minimum of Doppler factor because $\left|\alpha_{+} \right| \gg 1/\gamma$. Then, the maximum of intensity is reached for the second solution $\alpha_{-}$. The apparent velocity computed at this position is :
\begin{equation}
\beta_{\rm app} = \frac{\beta \sin \left | \alpha_{-}\right |}{1-\beta \cos \alpha_{-} } \simeq \frac{2 \gamma^2 \left |\alpha_{-}\right |}{1+\gamma^2 \alpha_{-}^2}
\end{equation}

Now if  $\left | \alpha_{-}\right |<1/\gamma$, and using Eq. (\ref{eq:Alphapm}), this reduces to :
\begin{equation}
\beta_{\rm app} \simeq \frac{\left | \dot{\gamma}\right |}{\gamma} \simeq \Delta \alpha^{-1}
\end{equation}

We can also check that an absence of velocity gradient in the jet ($\dot{\gamma}=0$) implies that the maximum of intensity is aligned with the line of sight ($\alpha_{-}=0$), resulting in the absence of apparent velocity.
This is a very interesting conclusion: for a smooth angular variation of the Lorentz factor, the apparent velocity of the jet is linked to the width of the angular velocity profile, and not on the absolute value of the maximal Lorentz factor.\\ 

In the case of a sharp variation, defined by  $\left | \dot{\gamma}\right |\gg\gamma^2$, the linearization of the square root yields a different result. We obtain two symmetrical solutions, independent of the velocity gradient:
\begin{equation}
\alpha_{\pm} \approx \pm\frac{1}{\gamma}
\end{equation}
This applies to the case of the conical jet profile $D_{1}$, for which there is a discontinuity in the velocity distribution. This discontinuity implies a infinite gradient at the edge of the jet. We can check on Fig.~\ref{fig:BetaAppProfil} that the maximal jet apparent velocity ($\beta_{\rmn{app}}=\gamma_{0}$) is reached for  $\alpha=1/\gamma$ i.e. for an observation angle $\theta_{\rmn{obs}}=\theta_{j}+1/\gamma$. In this case, the range of apparent velocities is the same as for a point source (see Fig. \ref{fig:BetaAppProfil}) as discussed in Sect. \ref{con_jet}. 
\begin{figure*}
\begin{tabular}{c}
\includegraphics [width=0.7\linewidth]{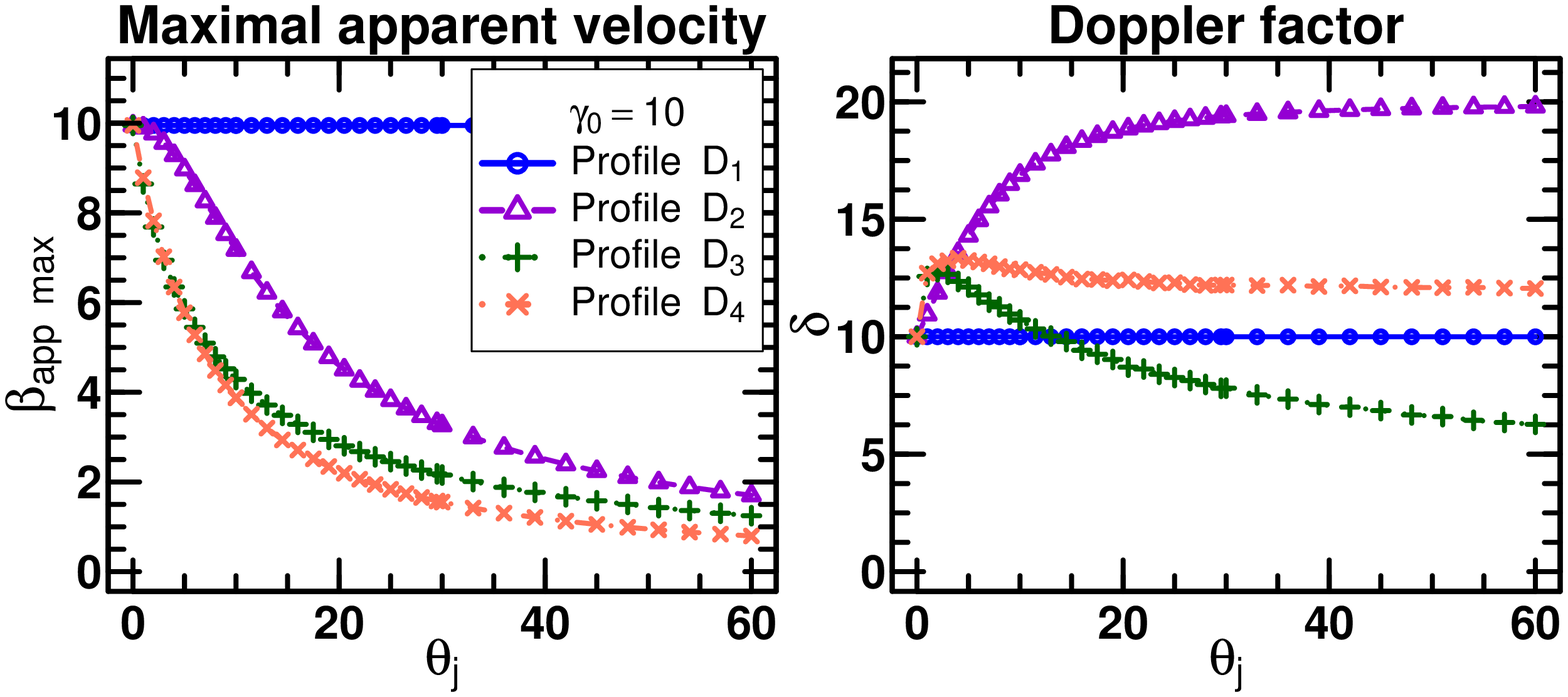}\\
\includegraphics [width=0.7\linewidth]{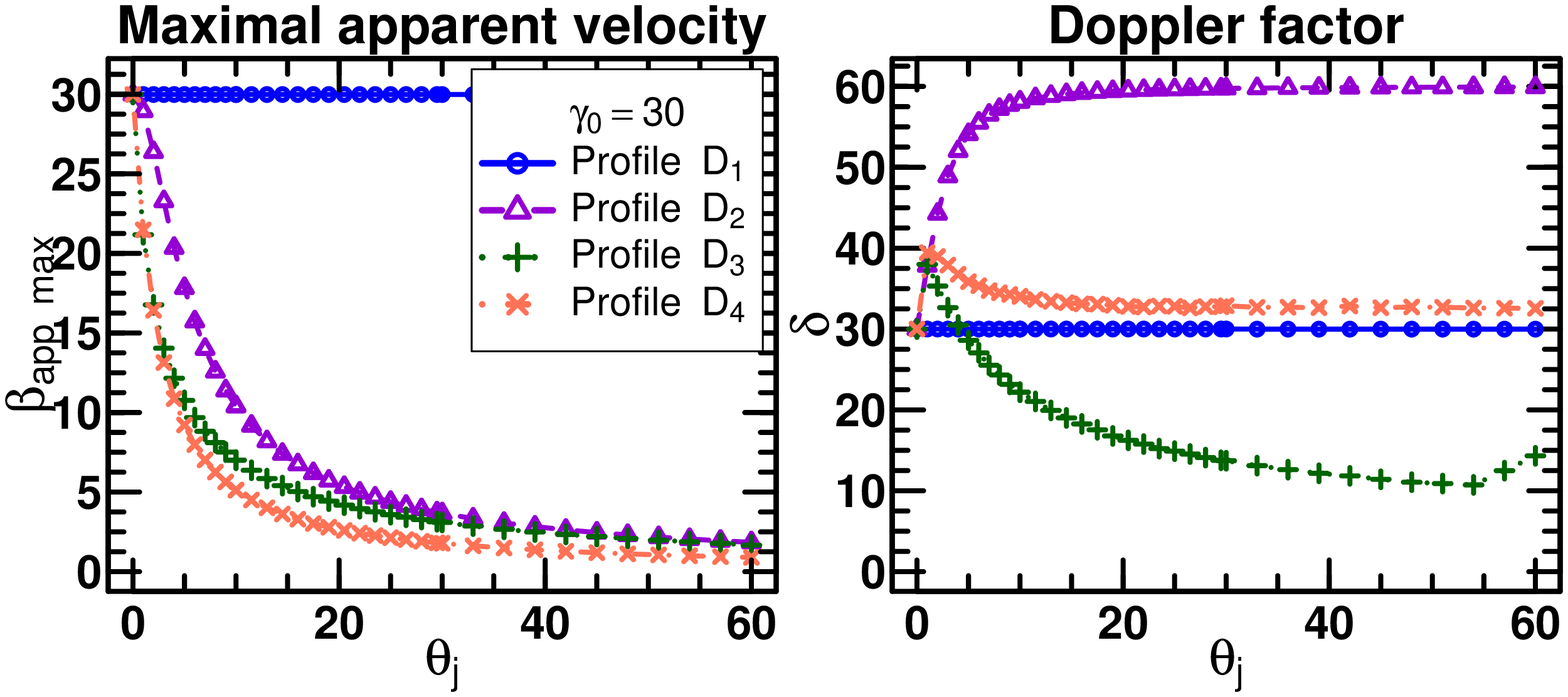}
\end{tabular}
\caption{{\bf Left:} Evolution of the maximal apparent velocity ($\beta_{\rmn{app,\,max}}$) as a function of the jet opening angle for the different velocity profiles $D_{1-4}$. Blue circles: profile $D_{1}$, violet triangles: profile $D_{2}$, green crosses: profile $D_{3}$, salmon crosses: profile $D_{4}$. For all profiles, we have set $\gamma_{0}=10$ (top) or $\gamma_{0}=30$ (bottom). {\bf Right:} Same as left, but for the Doppler factor.}
\label{fig:theta_j_var}
\end{figure*}
%
\subsection{Maximal apparent velocity}
\subsubsection{Effect of the jet opening angle}
For a given jet configuration ($D_{i}$, $\gamma_{0}$, $\theta_{j}$), we now define $\beta_{\rmn{app,\,max}}$, the {\it maximal} apparent jet velocity that can be obtained by varying the observation angle $\theta_{\rmn{obs}}$. In order to study the influence of the geometrical collimation, we have computed the evolution of $\beta_{\rmn{app,\,max}}$ for the different velocity profile $D_{1-4}$, as a function of the jet opening angle $\theta_{j}$. The results are presented in Fig.~\ref{fig:theta_j_var}, where we have plotted $\beta_{\rmn{app,\,max}}(\theta_{j})$ and the associated Doppler factor $\delta_{\beta_{app,max}}(\theta_{j})$ for each velocity profile and assuming a Lorentz factor $\gamma_{0}=10$ on the jet axis. \\

For the profile $D_{1}$ (conical profile), the maximal apparent velocity is constant,  $\beta_{\rmn{app,\,max}}\simeq\gamma_{0}$. The corresponding Doppler factor  $\delta_{\beta_{app,max}}$ is then of the order of $\beta_{\rmn{app,\,max}}$. As already mentioned, this profile can be seen as a point source when the observer looks outside the cone ($\theta_{\rmn{obs}}>\theta_{j}$). Then, it is always possible to find $\theta_{\rmn{obs}}$ that matches the conditions to obtain the maximal theoretical value $\beta_{\rmn{app}}=\gamma_{0}$.
 
The three other profiles $D_{2-4}$ have all a similar behavior which is however very different from the $D_1$ case. For small opening angles ($\theta_{j}\ll\gamma_{0}^{-1}$), $\beta_{\rmn{app,\,max}}$ is close to the theoretical maximum because in that case the jet is very well collimated and can be seen as a point-like source. When $\theta_j$ starts to increase, the value of $\beta_{\rmn{app,\,max}}$ decreases rapidly. Depending on the velocity profile, the corresponding Doppler factor increases or decreases, but remains much higher than the value of $\beta_{\rmn{app,\,max}}$. Hence, taking into account the effect of the jet opening angle allows to obtained small apparent velocities but with higher Doppler factors than in the homokinetic case.

%
\subsubsection{Empirical estimate of the jet collimation}
We have computed, for each velocity profile, surfaces of the maximum apparent velocity $\beta_{\rmn{app,\,max}}$ and the associated Doppler factor  $\delta_{\beta_{app,max}}$ in the ($\theta_{j},\gamma_{0}$) plane. Plots reported in Fig. \ref{fig:theta_j_var} then correspond to a section of these surfaces at a constant $\gamma_0$.

\begin{figure*}
\begin{tabular}{c}
\centerline{\hrulefill$\ $Profile $D_1$ \hrulefill}\\
\centerline{{ Maximal apparent velocity}\hspace*{2.5cm} { Corresponding Doppler factor}}\\
\includegraphics [width=0.7\textwidth]{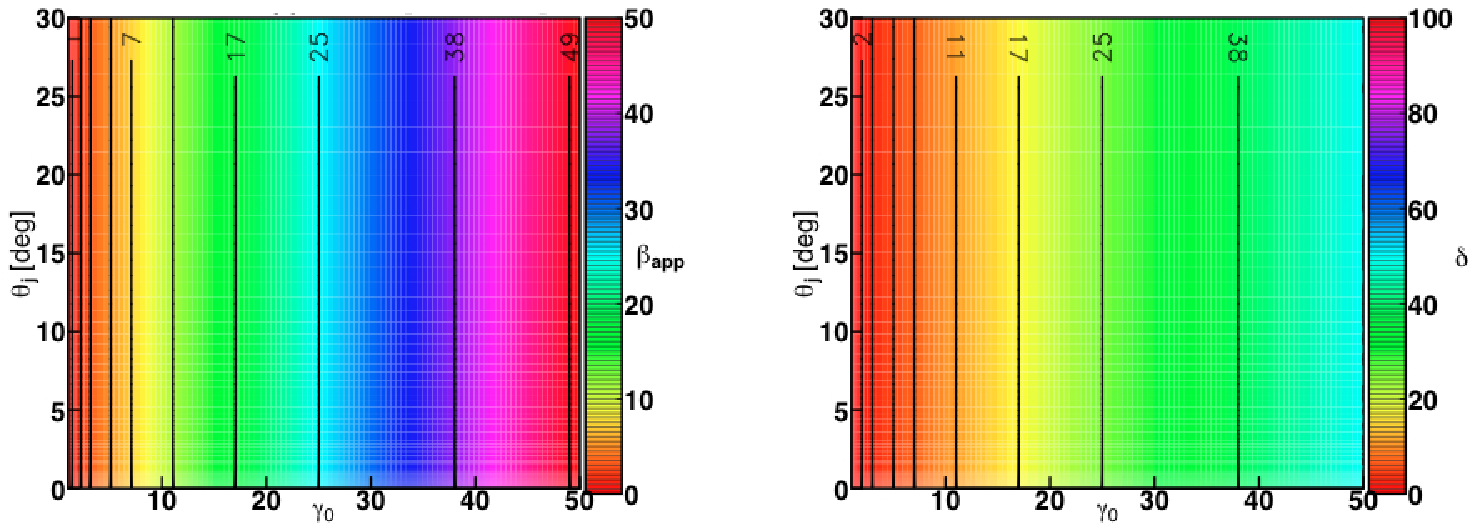}\\
\centerline{\hrulefill$\ $Profile $D_2$ \hrulefill}\\
\centerline{{ Maximal apparent velocity}\hspace*{2.5cm} { Corresponding Doppler factor}}\\
\includegraphics [width=0.7\textwidth]{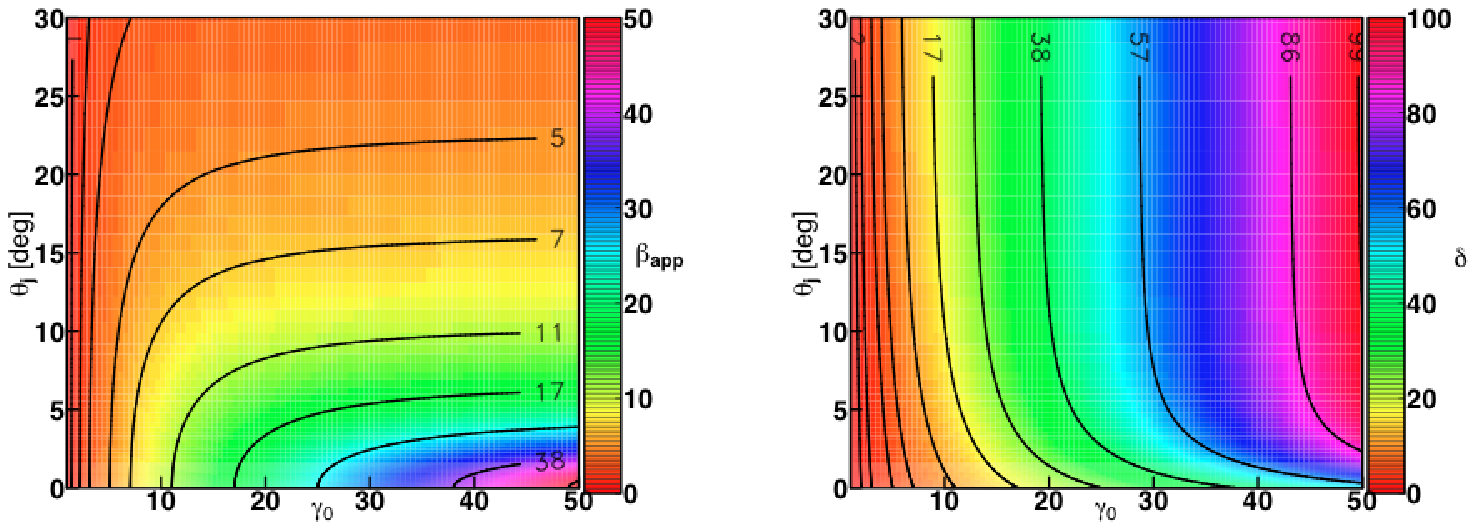}\\
\centerline{\hrulefill$\ $Profile $D_3$ \hrulefill}\\
\centerline{{ Maximal apparent velocity}\hspace*{2.5cm} { Corresponding Doppler factor}}\\
\includegraphics [width=0.7\textwidth]{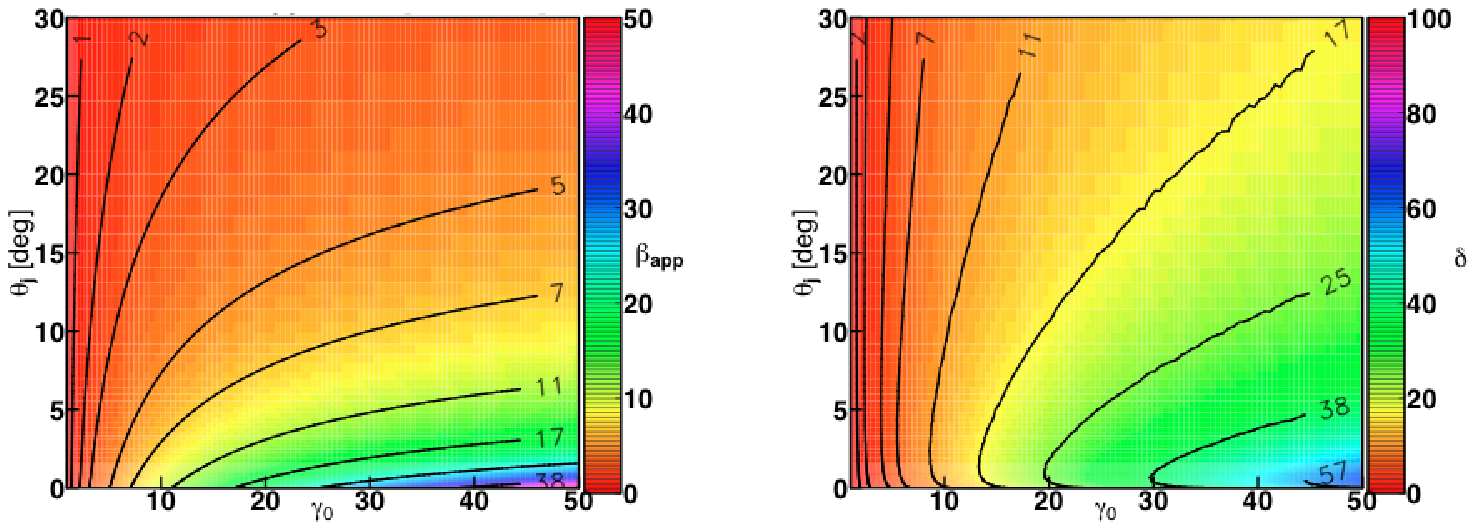}\\
\centerline{\hrulefill$\ $Profile $D_4$ \hrulefill}\\
\centerline{{ Maximal apparent velocity}\hspace*{2.5cm} { Corresponding Doppler factor}}\\
\includegraphics [width=0.7\textwidth]{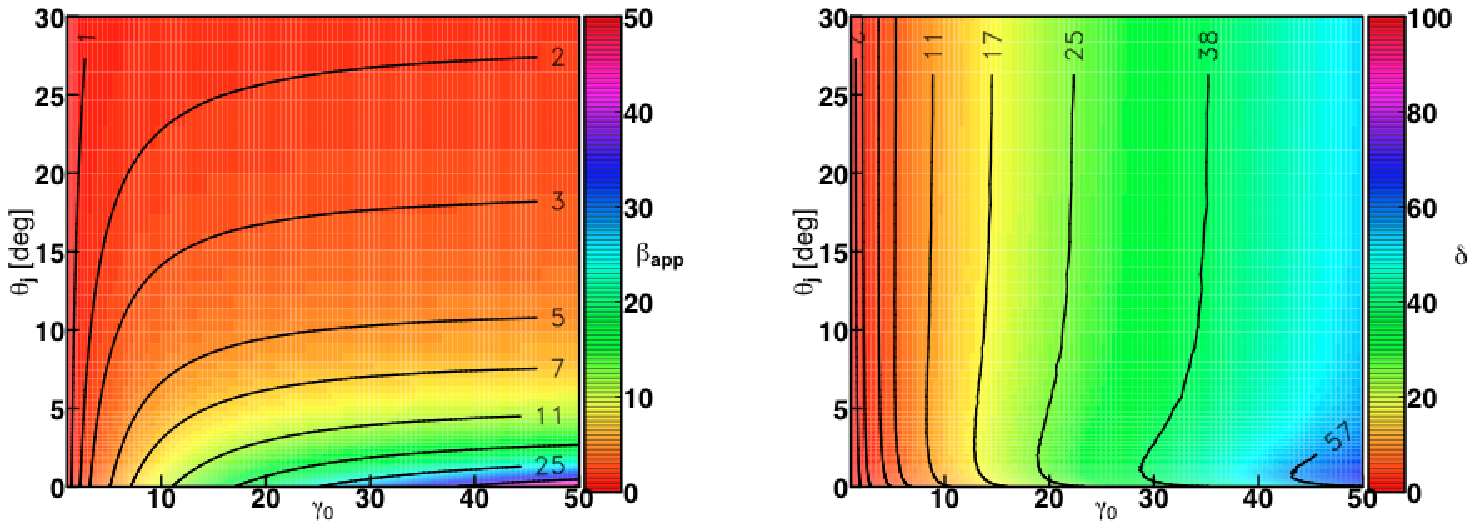}
\end{tabular}
\caption{Contour plots of the maximal apparent velocity ($\beta_{\rmn{app,\,max}}$ on the left) and the associated Doppler factor (on the right), as a function of the geometrical collimation of the jet $\theta_{j}$ and of the Lorentz factor on the jet axis $\gamma_{0}$, for the 4 different angular velocity profiles (from top to bottom) $D_{1}$,  $D_{2}$, $D_{3}$  and $D_{4}$ (Eq.~\ref{eq:D1}-\ref{eq:D4}).}
\label{fig:abaq}
\end{figure*}

These surfaces are plotted in Fig.~\ref{fig:abaq} with contours of $\beta_{\rmn{app,\,max}}$ and   $\delta_{\beta_{app,max}}$ in solid lines. In the case of the conical profile (top panel of Fig.~\ref{fig:abaq}), the maximum apparent velocity follows the theoretical expectations of a relativistic point source (see previous section) with  $\beta_{\rmn{app,\,max}}(\theta_{j},\gamma_{0})\simeq\gamma_0$ and a corresponding Doppler factor $\delta_{\beta_{app,max}}\simeq\gamma_0$, both parameters being independent of the jet opening angle $\theta_j$. The other velocity profiles give completely different dependencies on $\theta_j$ but also on $\gamma_0$ (see the 3 lower panels of Fig.~\ref{fig:abaq}), the maximum apparent velocity becoming strongly dependent on the jet opening angle, while the corresponding Doppler factor is very sensitive to the velocity profile. It confirms the affirmations that we make in the previous sections, i.e. that the jet opening angle decreases dramatically the apparent velocity, even for high jet Lorentz factor. And the larger the jet opening angle, the smaller the value of $\beta_{\rmn{app,\,max}}$ compared to the point-source case.\\

But more interestingly, Fig.~\ref{fig:abaq} allows to determine an empiric relation between the jet apparent velocity an its collimation angle for the continuous velocity profiles $D_{2-4}$: for instance, the collimation must be better than $\theta_{j} \la 5\,^\circ$, in order to observe an apparent velocity of at least $10c$ (i.e. $\beta_{{app,\,max}}>10$). This limit on the collimation can be expressed as $ 5\,^\circ \sim 0.1\,\rmn{rad}\sim \beta_ {{app,\,max}}^{-1}$ and can then  be  re-written as:
\begin{equation}
\theta_{j} \la \beta_{{app,\,max}}^{-1}.
\label{eqbetmax}
\end{equation}
This relation remains true by a factor of a few for the profiles $D_{2-4}$ { but appears to better work for continuous velocity profiles like $D_3$ and $D_4$. The abrupt variation of $\gamma$ in $D_2$ helps in reaching large apparent velocity for jet opening angle a bit larger than $\beta_{{app,\,max}}^{-1}$. However, since we expect continuous profiles to be more realistic, Eq. (\ref{eqbetmax}) should be relatively general.} Hence, the measurement of the apparent velocity of a relativistic jet can provide an upper limit on its collimation angle.

{
\subsection{Effect of an angular dependent emissivity}
\label{Iangdep}
In all the previous sections, we assumed a constant specific intensity in the shell frame, in all directions. The variation of apparent brightness is thus entirely due to varying Doppler factors. This is obviously a very crude assumption since the physical reasons causing a variation of jet velocity are also likely to produce variations of non thermal acceleration processes and thus of non thermal radiation emission. Considering a possible variation of local emissivity with the angle will add another complexity to the problem. We can remark however that for a given direction of the line of sight, a variable specific intensity can be considered as equivalent to a modification of the Doppler factor, since it is always possible to find another velocity producing the same apparent specific intensity at each position on the sky. Letting the specific intensity free to vary won't probably change much the qualitative conclusions about the limitation of apparent velocities and Doppler factors. \\

We illustrate the influence of a variable intensity with the $D_3$ distribution (gaussian profile) of the Lorenz factor and two different assumptions on the specific intensity.  The results are displayed in Fig. \ref{fig:VarIntensity} for $\gamma_0 =5$. We chose either a constant profile of the specific intensity (thick solid line) as computed previously, or a gaussian profile (thick dashed line). As already noticed, a constant intensity and Lorentz factor within a finite angle $\theta_j$ (conical profile) will always give a vanishing apparent velocity for $\theta_{obs}<\theta_j$ since the maximal Doppler factor will always been along the line of sight, giving a null projection on the sky. For $\theta_{obs}>\theta_j$ the image is dominated by the emission coming from the edge of the conical jet and the situation becomes similar to a point-like source. This is exemplified by the thin solid line in Fig. \ref{fig:VarIntensity}. Things are different with a variable Lorenz factor and variable intensity, since the maximum is no more strictly on the light of sight (it is determined by some compromise between the variation of the Doppler factor and the specific intensity). The apparent velocity will also be not zero, since it is directly related to the angular separation between the maximum and the line of sight. However, we note that the resulting pattern is not strongly modified by the inclusion of an angular dependency of the specific intensity profile and most of the effects are already catched by using an angle-dependent profile for the Lorenz factor. Since we don't have anyway a good knowledge of the real distribution of the Lorentz factors, the inclusion of a angular dependency of the emissivity wouldn't really add any new effects. 
 \begin{figure}
\includegraphics [height=50mm]{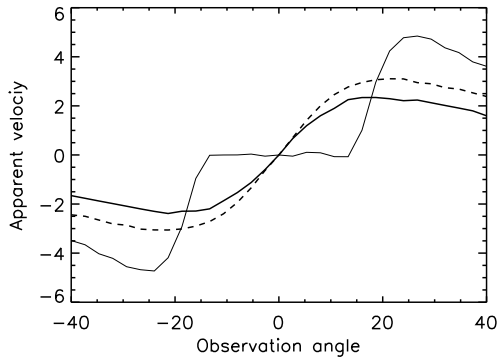}
\caption{Evolution of the apparent velocity as a function of the observation angle for a constant intensity profile (thick solid line) or a angular dependent (gaussian in the present case) intensity profile (thick dashed line). We use a gaussian profile ($D_3$) for the Lorenz factor. For comparison the case of a conical profile ($D_1$) and constant intensity is plotted in thin line. The jet opening angle $\theta_j=15\,^\circ$ and $\gamma_0=5$}
\label{fig:VarIntensity}
\end{figure}
}

%
\section{Discussion}\label{sec:disc}
All these simulations show that the apparent velocity of a jet is highly dependent on the geometrical collimation and the velocity profile inside the jet. The apparent velocity decreases dramatically compared to the homokinetic case as soon as the jet collimation angle is larger than a few degrees. These results are in agreement with those of \cite{gop04,gop06,gop07a,gop07b}, although the inclusion of some effects neglected by these authors modify the precise distribution. Our choice of another prescription to define the apparent velocity, as the velocity of the maximum of the specific intensity of the observed pattern, gives rise also to different values: for instance, as we showed in Sect. \ref{Appvel},  the number of objects having a null apparent velocity is finite in our case for a conical, single velocity jet, whereas it is true only exactly on-axis with the prescription of GK07 where the apparent superluminal velocity is obtained by averaging the different apparent velocities of the observed pattern. \\

As shown in Fig. \ref{fig:BetaAppProfil}, the distribution of apparent velocities is significantly different from that deduced from a single velocity Doppler boosted jet. This can strongly modify the statistical distribution of apparent velocities and thus the bulk Lorentz factors deduced from these statistical studies. 

In the case of a smooth decrease of the bulk Lorentz factor with the angle to the LOS, we have shown that the apparent velocities are much more indicative of the geometrical shapes of the jet than of the real bulk Lorentz factors. Thus the bulk Lorentz factors deduced from radio observations and statistical studies can be very different from those deduced from radiative constraints for high energy emission, and will give generally much lower values. This can obviously help solving the discrepancies between the two estimates, since Lorentz factors deduced from gamma-rays transparency arguments routinely exceeds several tens for TeV blazars,  although superluminal motion are seldom observed ({ e.g. \citealt{edw02,pin04,tav98,sau04}}). Note however that the discrepancy can be partly reduced with inhomogeneous models of jets (e.g. \citealt{hen06}). 

 A sharp decrease of the Lorentz factor outside a cone of almost constant Lorentz factors will give a different appearance: in this case, the apparent superluminal motion will be very small, maybe undetectable, inside the collimation cone, although the Doppler factor is maximal. But a small fraction of  objects, seen at a small angle from the edge of the cone, can exhibit large superluminal motions. Both zero apparent velocities and highest apparent velocities will be overrepresented in statistical samples compared to the homokinetic hypothesis. These possibilities should of course be carefully taken into account in statistical studies.\\ 
 
Contrary to apparent superluminal velocities, Fig.~\ref{fig:abaq} shows that the effective Doppler factor strongly depends on the precise velocity profile inside the jet. This can be understood by remarking that, in the case of a relativistic point source (Fig. \ref{fig:BetaAppTest}), when the apparent velocity is maximal, near $\theta_{obs} \simeq \gamma_0^{-1}$, the Doppler factor is varying very rapidly. Thus, small differences in the precise angle of the emitting regions contributing the most to the intensity can lead to significative differences in the Doppler factor, even if the apparent velocities are comparable. For instance, increasing the opening angle of the jet will generally decrease the apparent velocity, but can decrease or increase the effective Doppler factor (see Fig. \ref{fig:theta_j_var}). 

In consequence, in the case of open jets, it is difficult to derive a statistics of Doppler factors from the observation of apparent superluminal motions without knowing the precise angular distribution of Lorentz factors. \\
 
These considerations may give a hint to solve several discrepancies that arise when comparing superluminal motions and other constraints derived from radiative models or statistical arguments. Namely many TeV blazars show little or no apparent motion, whereas radiative constraints seem to be compatible only with very high bulk Lorentz factors. As most of TeV blazars are BL Lacs, which are thought to be beamed counterparts of weakly collimated FRI galaxies, it is tempting to explain this discrepancy by a rather large opening angle. It is interesting to note that very open jets are commonly invoked in the context of gamma-ray bursts, where the usually assumed bulk Lorentz factors are much higher (a few hundreds) than $\theta_j^{-1}$  (see however the "cannonball" model, \citealt{dar00}). The visual appearance of a gamma-ray burst in the earlier phase would be thus dominated by the part of the jet traveling at zero angle with respect to the line of sight, i.e. with no superluminal motion at all. \\
  
It should be stressed that we studied only a variation of $\gamma$ with the angle of ejection, but that other possibilities exist. One could also imagine that several (and may be a whole distribution) of Lorentz factors exist along a particular direction, even for a perfectly collimated jet. This is indeed necessary for all "shock in jets" models (see \citealt{zen97} for a review), where layers at different velocities are supposed to be ejected and collide through internal shocks. So called "spine in jets" and "two-flows" models \citep{ghi05,hen91} also assume different Lorentz factors along a single direction. We didn't consider either the possible variation of the bulk Lorentz factors with the distance. All these factors would of course complicate the picture, since they would produce different apparent velocities and bulk Lorentz factors along the jet, and most probably varying following the observed wavelengths. 

 \section{Conclusion}\label{sec:conc}
 We have studied the influence of the geometrical opening of relativistic jets on their apparent velocities and associated Doppler factors. We parametrized this opening by various distributions of Lorentz factor as a function of the angle of observation.  We have investigated the appearance of a thin shell ejected and traveling along the flow. We propose a new criterion to define the apparent velocity of the jet pattern as the velocity of the point having the highest observed specific intensity. The effective Doppler factor is then the Doppler factor associated with this point.\\ 
 
Studies of different configurations shows that the variation of the maximal apparent velocities as a function of the viewing angle can be strongly modified, especially when the Lorentz factor angular distribution is decreasing smoothly. In this case, the maximal apparent velocity is essentially determined by the typical angular interval on which the decrease occurs (which is also the opening angle of the jet for reasonable configurations), and not by the bulk Lorentz factor. The associated Doppler factor is a sensitive function of the precise shape of the distribution. This could help understanding the discrepancies between various estimates of bulk Lorentz factors in extragalactic jets.\\ 
 
 These findings could have significant impacts on statistical studies trying to estimate Doppler factors from statistical comparisons between beamed and unbeamed objects, in the frame of so-called "unification" models (e.g. \citealt{urr95}). The opening of jets modifies strongly the angular distribution of observed effective Doppler factors, and thus the statistical distribution for randomly oriented jets. It could thus significantly alter the conclusions of such studies. Further work will be done to investigate this issue. 

\section*{Acknowledgments}
{The authors acknowledge financial support from the GDR PCHE in France and the CNES french spatial agency}


\label{lastpage}
\end{document}